\documentclass[english,prd,superscriptaddress,nofootinbib,preprintnumbers]{revtex4} 

\usepackage{graphicx}
\usepackage{dcolumn}
\usepackage{bm}
\usepackage{latexsym}
\usepackage{amsfonts}
\usepackage{amssymb}
\usepackage{amsmath}
\usepackage{bm}
\usepackage{color}

\def\be{\begin{equation}}
\def\ee{\end{equation}}
\def\ba{\begin{eqnarray}}
\def\ea{\end{eqnarray}}

\begin{document}

\preprint{KOBE-TH-13-09}

\title{Anisotropic power-law k-inflation}

\author{Junko Ohashi}

\affiliation{Department of Physics, Faculty of Science, Tokyo University of Science, 
1-3, Kagurazaka, Shinjuku, Tokyo 162-8601, Japan}

\author{Jiro Soda}

\affiliation{Department of Physics, Kobe University, Kobe 657-8501, Japan}

\author{Shinji Tsujikawa}

\affiliation{Department of Physics, Faculty of Science, Tokyo University of Science, 
1-3, Kagurazaka, Shinjuku, Tokyo 162-8601, Japan}

\date{\today}


\begin{abstract}

It is known that power-law k-inflation can be realized for the Lagrangian 
$P=Xg(Y)$, where $X=-(\partial \phi)^2/2$ is the kinetic energy 
of a scalar field $\phi$ and $g$ is an arbitrary function in terms of  
$Y=Xe^{\lambda \phi/M_{\rm pl}}$ 
($\lambda$ is a constant and $M_{\rm pl}$ is the reduced Planck mass). 
In the presence of a vector field coupled to the inflaton with 
an exponential coupling $f(\phi) \propto e^{\mu \phi/M_{\rm pl}}$, 
we show that the models with the Lagrangian $P=Xg(Y)$ generally 
give rise to anisotropic inflationary solutions with 
$\Sigma/H={\rm constant}$, where $\Sigma$ is an anisotropic 
shear and $H$ is an isotropic expansion rate. 
Provided these anisotropic solutions exist in the regime 
where the ratio $\Sigma/H$ is much smaller than 1, 
they are stable attractors irrespective of the 
forms of $g(Y)$. We apply our results to 
concrete models of k-inflation such as the generalized
dilatonic ghost condensate/the DBI model and we numerically show 
that the solutions with different initial conditions converge 
to the anisotropic power-law inflationary attractors. 
Even in the de Sitter limit ($\lambda \to 0$) such solutions
can exist, but in this case the null energy condition is generally violated. 
The latter property is consistent with the Wald's
cosmic conjecture stating that the anisotropic hair does not 
survive on the de Sitter background in the presence of matter 
respecting the dominant/strong energy conditions. 

\end{abstract}


\maketitle


\section{Introduction}

The inflationary paradigm, which was originally proposed in \cite{oldinf},  
is now widely accepted as a viable phenomenology describing the cosmic acceleration 
in the very early Universe. The simplest inflationary scenario based on 
a single scalar field predicts the generation of nearly scale-invariant and 
adiabatic density perturbations \cite{oldper}. This prediction is in agreement with 
the temperature fluctuations of the Cosmic Microwave Background (CMB) observed 
by the WMAP \cite{WMAP} and Planck \cite{Planck} satellites.

The WMAP data showed that there is an anomaly associated with the 
broken rotational invariance of the CMB perturbations \cite{aniobser}. 
This implies that the statistical isotropy of the power spectrum of 
curvature perturbations is broken, which is difficult to be addressed 
in the context of the simplest single-field inflationary scenario. 
Although we cannot exclude the possibility that some systematic
effects cause this anisotropy \cite{Hanson}, it is worth exploring  
the primordial origin of such a broken rotational invariance.

If the inflaton field $\phi$ couples to a vector kinetic term $F_{\mu \nu}F^{\mu \nu}$, 
an anisotropic hair can survive during inflation for a suitable choice 
of the coupling $f^2(\phi)$ \cite{Watanabe}. 
In such cases, the presence of the vector field 
gives rise to the anisotropic power spectrum 
consistent with the broken rotational invariance of 
the CMB perturbations \cite{Gum,Watanabe:2010fh} 
(see also Refs.~\cite{Yoko}-\cite{Rodriguez} for related works).
In addition, the models predict the detectable level of non-Gaussianities 
for the local shape averaged over all directions with respect to 
a squeezed wave number \cite{Bartolo,Shiraishi}. 
In the two-form field models where the inflaton 
couples to the kinetic term $H_{\mu \nu \lambda}H^{\mu \nu \lambda}$
the anisotropic hair can also survive \cite{Ohashi}, but their observational signatures
imprinted in CMB are different from those in the vector model \cite{Ohashi2}.

For a canonical inflaton field with the potential $V(\phi)$, the 
energy density of a vector field can remain nearly constant for the coupling 
$f(\phi)=\exp[\int 2V/(M_{\rm pl}^2V_{,\phi})d\phi]$ \cite{Watanabe}, 
where $V_{,\phi}=dV/d\phi$.
For the exponential potential $V(\phi)=ce^{-\lambda \phi/M_{\rm pl}}$
the coupling is of the exponential form $f(\phi)=e^{-2\phi/(\lambda M_{\rm pl})}$, 
as it often appears in string theory and supergravity \cite{Ratra}. 
In this case there exists an anisotropic power-law inflationary attractor along which 
the ratio $\Sigma/H$ is constant \cite{Kanno10}, where $\Sigma$ is an anisotropic 
shear and $H$ is an isotropic expansion rate. 
For general slow-roll models in which the cosmic acceleration 
comes to end, the solution with an anisotropic hair corresponds to 
a temporal attractor during inflation \cite{Watanabe}.

There exists another inflationary scenario based on 
the scalar-field kinetic energy $X=-(\partial \phi)^2/2$ 
with the Lagrangian $P(\phi,X)$-- dubbed k-inflation \cite{kinf}. 
The representative models of k-inflation are the (dilatonic) ghost 
condensate \cite{Arkani,Piazza} and 
the Dirac-Born-Infeld (DBI) model \cite{DBI}. 
In such cases the evolution of the inflaton can be 
faster than that of the standard slow-roll inflation, so the coupling 
$f(\phi)$ with the vector field can vary more significantly.
It remains to see whether the anisotropic hair survives in k-inflation. 
This is important to show the generality of anisotropic inflation.

In Refs.~\cite{Piazza,Sami} it was found that in the presence of 
a scalar field and a barotropic perfect fluid the condition for the existence of scaling 
solutions restricts the Lagrangian of the form $P(\phi,X)=Xg(Y)$, where $g$ is an 
arbitrary function in terms of $Y=X e^{\lambda \phi/M_{\rm pl}}$ 
and $\lambda$ is a constant. 
On the flat Friedmann-Lema\^{i}tre-Robertson-Walker (FLRW) background 
there exists a scalar-field dominated attractor responsible for inflation 
under the condition $\lambda^2<2\,\partial P/\partial X$ \cite{Tsuji06,Ohashi2011}.
In fact, the Lagrangian $P(\phi,X)=Xg(Y)$ covers a wide class of 
power-law inflationary scenarios such as the canonical scalar field 
with the exponential potential ($g(Y)=1-cM_{\rm pl}^4/Y$), the dilatonic ghost 
condensate ($g(Y)=-1+cY/M_{\rm pl}^4$), and the DBI 
model ($g(Y)=-(m^4/Y)\sqrt{1-2Y/m^4}-M^4/Y$).
There is also another power-law inflationary scenario studied in Ref.~\cite{Unnikrishnan:2013vga}.

In the presence of a vector kinetic term $F_{\mu \nu}F^{\mu \nu}$
with the coupling $f(\phi)=f_0e^{-\mu \phi/M_{\rm pl}}$, the canonical scalar 
field with the exponential potential $V(\phi)=ce^{-\lambda \phi/M_{\rm pl}}$
gives rise to stable anisotropic inflationary solutions under the 
condition $\lambda^2+2\mu \lambda-4>0$ \cite{Kanno10}. 
For the power-law DBI inflation it was shown that the anisotropic 
hair can survive under certain conditions \cite{Kao}
(see also Ref.~\cite{Tachyon} for the power-law tachyon inflation).
In this paper we study the existence and the stability of anisotropic 
fixed points for the general Lagrangian $P(\phi,X)=Xg(Y)$.
Remarkably, if anisotropic inflationary fixed points exist, they are 
stable irrespective of the forms of $g(Y)$ in the regime 
where the anisotropy is small ($\Sigma/H \ll 1$).

This paper is organized as follows. 
In Sec.~\ref{backsec} we derive the equations of motion for the Lagrangian 
$P(\phi,X)$ on the anisotropic cosmological background.
In Sec.~\ref{fixedsec} we obtain anisotropic fixed points for the 
Lagrangian $P=Xg(Y)$ and discuss the stability of them against 
the homogenous perturbations.
In Sec.~\ref{concretesec} we apply our general results to concrete 
models of power-law inflation and numerically confirm the existence 
of stable anisotropic solutions.
Sec.~\ref{consec} is devoted to conclusions.


\section{Background equations of motion}
\label{backsec}

Let us consider the theories described by the action 
\begin{equation}
S=\int d^4x\sqrt{-g_M}\left[ \frac{M_{\rm pl}^2}{2}R
+P(\phi,X)-\frac{1}{4} f(\phi )^2 F_{\mu\nu}F^{\mu\nu}  
\right]\,,
\label{eq:action}
\end{equation}
where $g_M$ is the determinant of the metric $g_{\mu \nu}$, 
$R$ is the scalar curvature, and $P(\phi, X)$ is a function with 
respect to the inflaton $\phi$ and its derivative 
$X=-(1/2)g^{\mu \nu} \partial_{\mu} \phi \partial_{\nu} \phi$. 
The field $\phi$ couples to a vector kinetic term 
$F_{\mu\nu}F^{\mu\nu}$, where the vector field $A_{\mu}$
is related to $F_{\mu\nu}$ as $F_{\mu\nu} = 
\partial_\mu A_\nu - \partial_\nu A_\mu$.

Choosing the gauge $A_0=0$, we can take the $x$-axis for 
the direction of the vector field, i.e., 
$A_{\mu}=(0,v(t),0,0)$, where $v(t)$ is a function of the cosmic time $t$.
Since there is the rotational symmetry in the $(y,z)$ plane, 
we take the line element of the form
\begin{equation}
ds^2 = -{\cal N}(t)^2 dt^2 + e^{2\alpha(t)} \left[ e^{-4\sigma (t)}dx^2
+e^{2\sigma (t)}(dy^2+dz^2) \right] \ ,
\label{anisotropic-metric}
\end{equation}
where ${\cal N}(t)$ is the Lapse function, 
$e^\alpha \equiv a$ and $\sigma$ are the isotropic 
scale factor and the spatial shear, respectively.
For this metric the action (\ref{eq:action}) reads
\begin{equation}
S=\int d^4 x \frac{e^{3\alpha}}{{\cal N}} \left[ 3M_{\rm pl}^2
(\dot{\sigma}^2-\dot{\alpha}^2)+{\cal N}^2 P(\phi,X({\cal N}))
+\frac12 f(\phi)^2 e^{-2\alpha+4\sigma} \dot{v}^2 \right]\,,
\label{actioncon}
\end{equation}
where a dot represents a derivative with respect to $t$, 
and $X({\cal N})=\dot{\phi}^2 {\cal N}^{-2}/2$. 
The field equation of motion for the field $v$ following from 
the action (\ref{actioncon}) is integrated to give
\begin{equation}
\dot{v}=p_A\,f(\phi)^{-2} e^{-\alpha-4\sigma}\,,
\end{equation}
where $p_A$ is an integration constant.
Varying the action (\ref{actioncon}) with respect to 
${\cal N}$, $\alpha$, $\sigma$, $\phi$, and setting 
${\cal N}=1$, it follows that  
\begin{eqnarray}
& &H^2=\dot{\sigma}^2+\frac{1}{3M_{\rm pl}^2}
\left[ 2XP_{,X}-P+\frac{p_A^2}{2}f(\phi)^{-2} e^{-4\alpha-4\sigma} \right]\,,
\label{be1} \\
& &\ddot{\alpha}=-3\dot{\sigma}^2-\frac{1}{M_{\rm pl}^2}
\left[ XP_{,X}+\frac{p_A^2}{3} f(\phi)^{-2} e^{-4\alpha-4\sigma} \right]\,,
\label{be2} \\
& & \ddot{\sigma}=-3\dot{\alpha} \dot{\sigma}+
\frac{p_A^2}{3M_{\rm pl}^2}f(\phi)^{-2} e^{-4\alpha-4\sigma}\,,
\label{be3} \\
& &(P_{,X}+2XP_{,XX})\ddot{\phi}+3P_{,X} \dot{\alpha}\dot{\phi} 
+P_{,X \phi} \dot{\phi}^2-P_{,\phi} 
-p_A^2f(\phi)^{-3}f_{,\phi} (\phi)e^{-4\alpha-4\sigma}=0\,,
\label{be4}
\end{eqnarray}
where $H\equiv \dot{\alpha}=\dot{a}/a$ is the Hubble expansion rate, 
and $P_{,X} \equiv \partial P/\partial X$ etc. 
We define the energy densities of the inflaton 
and the vector field, respectively, as
\begin{equation}
\rho_{\phi} \equiv 2XP_{,X}-P\,,\qquad
\rho_A \equiv \frac{p_A^2}{2} f(\phi)^{-2}e^{-4\alpha -4\sigma} \,.
\label{rhoA}
\end{equation} 

In order to sustain inflation, we require the condition $\rho_{\phi} \gg \rho_A$.
Since the shear term $\Sigma \equiv \dot{\sigma}$ should be suppressed 
relative to $H$, Eq.~(\ref{be1}) reads
\begin{equation}
H^2 \simeq \frac{\rho_{\phi}}{3M_{\rm pl}^2}\,.
\label{Friap}
\end{equation}
On using the slow-roll parameter $\epsilon \equiv -\dot{H}/H^2$, 
Eq.~(\ref{be2}) can be written as 
\begin{equation}
\epsilon=\frac{3\dot{\sigma}^2}{H^2}+\frac{XP_{,X}}{M_{\rm pl}^2 H^2}
+\frac{2\rho_A}{3M_{\rm pl}^2 H^2}\,.
\label{epdef}
\end{equation}
Each term on the r.h.s. of this equation needs to be much smaller than 
unity. In particular, if the contributions of the shear and the vector-field
energy density are negligible, Eq.~(\ref{epdef}) reduces to 
the standard relation $\epsilon \simeq XP_{,X}/(M_{\rm pl}^2 H^2)$
of k-inflation \cite{kinf}.

{}From Eq.~(\ref{be3}) the shear term obeys
\begin{equation}
\dot{\Sigma}=-3H \Sigma+\frac{2\rho_A}{3M_{\rm pl}^2}\,.
\end{equation}
If $\Sigma$ converges to a constant value, it follows that 
\begin{equation}
\frac{\Sigma}{H} \simeq \frac{2\rho_A}{3\rho_{\phi}}\,,
\end{equation}
where we used Eq.~(\ref{Friap}). 
If the evolution of $\rho_A$ is proportional to $\rho_{\phi}$, the ratio 
$\Sigma/H$ remains constant. 
This actually happens for anisotropic inflationary attractors 
discussed in the next section.


%
\section{Power-law k-inflation and the stability of anisotropic fixed points}
\label{fixedsec}

On the flat isotropic FLRW background the power-law k-inflation can be realized 
by the following general Lagrangian \cite{Tsuji06,Ohashi2011}
\begin{equation}
P(\phi,X)=X\,g(Y)\,,\qquad Y \equiv X e^{\lambda \phi/M_{\rm pl}}\,,
\label{scalinglag}
\end{equation}
where $g$ is an arbitrary function of $Y$, and $\lambda$
is a constant. Originally, the Lagrangian (\ref{scalinglag}) was derived
for the existence of scaling solutions in the presence of
a barotropic perfect fluid \cite{Piazza,Sami}.
Under the condition $\lambda^2<2P_{,X}$ there exists 
a power-law inflationary solution for any functions of $g(Y)$ \cite{Tsuji06}.

For the choice $g(Y)=1-cM_{\rm pl}^4/Y$, where $c$ is a constant, 
the Lagrangian (\ref{scalinglag}) reduces to 
$P=X-c M_{\rm pl}^4 e^{-\lambda \phi/M_{\rm pl}}$ \cite{expo}, 
in which case the dynamics of anisotropic inflation was studied 
in Ref.~\cite{Kanno10}. 
The dilatonic ghost condensate model 
$P=-X+ce^{\lambda \phi/M_{\rm pl}}X^2/M_{\rm pl}^4$ \cite{Piazza} 
corresponds to the choice $g(Y)=-1+cY/M_{\rm pl}^4$. 
If we choose the function $g(Y)=-(m^4/Y)\sqrt{1-2Y/m^4}-M^4/Y$, 
we recover the DBI Lagrangian $P=-h(\phi)^{-1} \sqrt{1-2h(\phi)X}
+h(\phi)^{-1}-V(\phi)$ with $h(\phi)^{-1} =
m^4e^{-\lambda \phi/M_{\rm pl}}$ and 
$V(\phi)=(M^4+m^4)e^{-\lambda \phi/M_{\rm pl}}$.

In the following we study inflationary solutions for the Lagrangian 
(\ref{scalinglag}) on the anisotropic background given 
by the metric (\ref{anisotropic-metric}).

\subsection{Anisotropic fixed points}

For the Lagrangian (\ref{scalinglag}) the field equation of motion 
(\ref{be4}) reads
\begin{equation}
\ddot{\phi}+3HA(Y)P_{,X}(Y) \dot{\phi}+\frac{\lambda X}{M_{\rm pl}}
\left\{ 1-[g(Y)+2g_1(Y)]A(Y) \right\}-2 \frac{f_{,\phi}}{f} \rho_A A(Y)=0\,,
\label{be5}
\end{equation}
where 
\begin{equation}
g_n(Y)=Y^n \frac{dg^n(Y)}{dY^n}\,,\qquad 
P_{,X}(Y)=g(Y)+g_1(Y)\,,\qquad
A(Y)=[g(Y)+5g_1(Y)+2g_2(Y)]^{-1}\,.
\end{equation}
The quantity $A=(P_{,X}+2XP_{,XX})^{-1}$ is related to 
the sound speed $c_s$, as $c_s^2=P_{,X}A$ \cite{Garriga,Piazza}.

In order to study the dynamics of anisotropic power-law k-inflation, 
it is convenient to introduce 
the following dimensionless variables
\begin{equation}
x_1=\frac{\dot{\phi}}{\sqrt{6}HM_{\rm pl}}\,,\qquad
x_2=\frac{M_{\rm pl} e^{-\frac{\lambda \phi}{2M_{\rm pl}}}}{\sqrt{3}H}\,,
\qquad
x_3=\frac{\dot{\sigma}}{H}\,,\qquad
x_4=\frac{\sqrt{\rho_A}}{\sqrt{3}HM_{\rm pl}}\,.
\label{dimendef}
\end{equation}
The variable $Y$ is related to $x_1$ and $x_2$ via
\begin{equation}
Y/M_{\rm pl}^4=x_1^2/x_2^2\,.
\end{equation}
{}From Eq.~(\ref{be1}) there is the constraint equation 
\begin{equation}
x_4^2=1-x_3^2-x_1^2 (P_{,X}+g_1)\,,
\label{x4re}
\end{equation}
whereas Eq.~(\ref{be2}) gives
\begin{equation}
\frac{\dot{H}}{H^2}=-2-x_3^2-x_1^2 (P_{,X}-2g_1)\,.
\label{dotH}
\end{equation}
On using Eqs.~(\ref{be3}), (\ref{be5}), (\ref{x4re}), and (\ref{dotH}), 
we obtain the following autonomous equations
\begin{eqnarray}
\hspace{-0.3cm}
x_1'(N) &=& \frac12 x_1 \left[ 4+2x_3^2-\sqrt{6}\lambda x_1
+2x_1^2 (P_{,X}-2g_1)\right]
-\frac{\sqrt{6}A}{2}
\left[ \sqrt{6}x_1 P_{,X}-x_1^2 (P_{,X}+g_1)(\lambda+2\mu)
+2(1-x_3^2)\mu \right],\label{auto1} \\
\hspace{-0.3cm}
x_2'(N) &=& \frac12 x_2 \left[ 4+2x_3^2-\sqrt{6}\lambda x_1 
+2x_1^2 (P_{,X}-2g_1) \right]\,,
\label{auto2}\\
\hspace{-0.3cm}
x_3'(N) &=& (x_3^2-1)(x_3-2)+x_1^2 \left[ P_{,X} (x_3-2)
-2g_1(x_3+1) \right]\,,
\label{auto3}
\end{eqnarray}
where $N=\ln a$, $x_i'(N)=dx_i(N)/dN$ ($i=1,2,3$), and 
$\mu=-M_{\rm pl}f_{,\phi}/f$. In the following we focus on 
the case of constant $\mu$, i.e., the coupling 
\begin{equation}
f(\phi)=f_0 e^{-\mu \phi/M_{\rm pl}}\,,
\label{fphi}
\end{equation}
where $f_0$ is a constant.

The fixed points responsible for the cosmic acceleration 
correspond to non-zero values of $x_1$ and $x_2$.
Setting the r.h.s. of Eqs.~(\ref{auto1})-(\ref{auto3}) to be 0, 
we obtain the following two fixed points
\begin{itemize}
\item (i) Isotropic fixed point
\begin{equation}
P_{,X}(Y)=\frac{\lambda}{\sqrt{6}x_1}\,,\qquad
g_1(Y)=\frac{6-\sqrt{6}\lambda x_1}{6x_1^2}\,,\qquad
x_3=0\,,\qquad x_4=0\,.
\label{auiso} 
\end{equation}
\item (ii) Anisotropic fixed point
\begin{eqnarray}
P_{,X} (Y) &=& \frac{(\lambda+2\mu)[2\sqrt{6}-(\lambda+6\mu)x_1]}{8x_1}\,,
\qquad
g_1(Y) = \frac{[2\sqrt{6}-(\lambda+2\mu)x_1](\sqrt{6}-\lambda x_1)}
{8x_1^2}\,, \nonumber \\
x_3 &=& \frac{\sqrt{6}}{4} (\lambda+2\mu)x_1-1\,,
\qquad
x_4^2 = \frac18 [ 3(\lambda+2\mu)x_1-2\sqrt{6}]
(\sqrt{6}-\lambda x_1)\,.
\label{auani1} 
\end{eqnarray}
\end{itemize}

Provided $g(Y)$ is given, the quantities $Y$ and $x_1$ are known 
by solving the first two equations of (\ref{auiso}) or (\ref{auani1}). 
In the Appendix we discuss more explicit expressions of isotropic 
and anisotropic solutions corresponding to the fixed points 
(\ref{auiso}) and (\ref{auani1}), respectively.
For both the isotropic and anisotropic fixed points, 
the slow-roll parameter is simply given by 
\begin{equation}
\epsilon=-\frac{\dot{H}}{H^2}
=\frac{\sqrt{6}}{2}\lambda x_1\,,
\label{dotH3}
\end{equation}
where we used Eq.~(\ref{dotH}). 
If $\lambda x_1>0$, the power-law inflation 
$a \propto t^{2/(\sqrt{6}\lambda x_1)}$ is realized.
Violation of the condition $\lambda x_1>0$ means that the fixed 
points correspond to super-inflationary solutions with $\dot{H}>0$. 
Then, the condition for the cosmic acceleration with a decreasing 
Hubble parameter is given by 
\begin{equation}
0<\lambda x_1< \frac{\sqrt{6}}{3}\,.
\label{con1}
\end{equation}
The presence of the anisotropic fixed point (ii) implies that
$x_4^2>0$. This translates to
\begin{equation}
3(\lambda+2\mu)x_1>2\sqrt{6}\,,
\label{con2}
\end{equation}
where we used (\ref{con1}). 
Under the condition (\ref{con2}) we also have $x_3>0$. 

In the absence of the vector field coupled to $\phi$, the ghost 
is absent for $P_{,X}>0$. 
For the anisotropic fixed point (ii), the condition $P_{,X}>0$ 
corresponds to
\begin{equation}
(\lambda+6\mu)x_1<2\sqrt{6}\,,
\label{con3}
\end{equation}
where we employed the fact that, from Eq.~(\ref{con2}), 
the signs of $x_1$ and $\lambda+2\mu$ are the same.

{}From Eqs.~(\ref{be1}) and (\ref{be2}) the total energy density and pressure 
are given by $\rho_t=2XP_{,X}-P+p_A^2f(\phi)^{-2}e^{-4\alpha -4\sigma} /2$ 
and $P_t=P+p_A^2f(\phi)^{-2}e^{-4\alpha -4\sigma} /6$, respectively.
Then we have
\begin{equation}
\rho_t+P_t = 
2H^2M_{\rm pl}^2 (3x_1^2 P_{,X}+2x_4^2)\,. 
\label{nec}
\end{equation}
If $P_{,X}>0$, then the null energy condition (NEC) $\rho_t+P_t>0$ 
is automatically satisfied. At the anisotropic fixed point (ii),
it is also possible to satisfy the NEC even for $P_{,X}<0$.
Substituting Eq.~(\ref{auani1}) into Eq.~(\ref{nec}), it follows that 
\begin{equation}
\rho_t+P_t=
2H^2M_{\rm pl}^2 \left[ \sqrt{6} (3\mu+2\lambda)
x_1-9(\lambda+2\mu)^2x_1^2/8-3 \right]\,.
\label{nec2}
\end{equation}
Then, the NEC translates to 
\begin{equation}
\frac{2\sqrt{6}}{9} \frac{4\lambda+6\mu-\sqrt{\lambda(7\lambda+12\mu)}}
{(\lambda+2\mu)^2}<x_1<
\frac{2\sqrt{6}}{9} \frac{4\lambda+6\mu+\sqrt{\lambda(7\lambda+12\mu)}}
{(\lambda+2\mu)^2}\,,
\label{x1upper}
\end{equation}
whose existence requires that $\lambda(7\lambda+12\mu)>0$.
Let us consider the case where $\lambda>0$ and $\mu>0$.
As long as the upper bound of Eq.~(\ref{x1upper}) is larger than 
the value $2\sqrt{6}/[3(\lambda+2\mu)]$, there are some values of 
$x_1$ consistent with both (\ref{con2}) and (\ref{x1upper}). 
This is interpreted as the condition 
$\lambda+\sqrt{\lambda (7\lambda+12\mu)}>0$, 
which is in fact satisfied for $\lambda>0$. 

In the limit that $\lambda \to 0$ the condition (\ref{con2}) 
reduces to $x_1>\sqrt{6}/(3\mu)$, while the region (\ref{x1upper}) shrinks 
to the point $x_1=\sqrt{6}/(3\mu)$.
When $\lambda=0$, Eq.~(\ref{nec2}) reads
\begin{equation}
\rho_t+P_t=-H^2 M_{\rm pl}^2 ( \sqrt{6}-3\mu x_1)^2\,,
\end{equation}
which is negative for $x_1>\sqrt{6}/(3\mu)$. 
Notice that, from Eq.~(\ref{dotH3}), the limit $\lambda \to 0$ 
corresponds to the de Sitter solution with constant $H$.
Hence the NEC is generally violated on the de Sitter solution.
The violation of the NEC means that the dominant energy condition 
(DEC; $\rho_t \geq |P_t|$) as well as the strong energy 
condition (SEC; $\rho_t+P_t \geq 0$ and $\rho_t+3P_t \geq 0$) 
are not satisfied \cite{Carroll}. 
This property is consistent with the Wald's cosmic no-hair 
conjecture \cite{Wald} stating that, in the presence of 
an energy-momentum tensor satisfying both DEC and SEC, 
the anisotropic hair does not survive on the de Sitter background.

In summary, for $\lambda>0$ and $\mu>0$, the anisotropic fixed points 
satisfying both $P_{,X}>0$ and the NEC exist in the regime
\begin{equation}
\frac{2\sqrt{6}}{3(\lambda+2\mu)}<x_1<\frac{2\sqrt{6}}
{\lambda+6\mu}\,,
\label{x1confin}
\end{equation}
whose upper bound (which comes from $P_{,X}>0$) gives a tighter constraint 
than that in Eq.~(\ref{x1upper}) (which comes from the NEC). 
As long as $\lambda>0$, there are some allowed values of $x_1$ which 
exist in the region (\ref{x1confin}).
Under the condition (\ref{x1confin}) the anisotropic parameter 
$x_3=\Sigma/H$ is in the range
\begin{equation}
0<x_3<\frac{2}{1+6\mu/\lambda}\,,
\label{x3upper}
\end{equation}
whose upper limit is determined by the ratio $\mu/\lambda$.
For compatibility of the two conditions (\ref{con1}) and (\ref{con2})
we require that $\mu/\lambda>1/2$. 
Hence the anisotropic parameter is generally constrained 
to be $x_3<1/2$.

\subsection{Stability of the anisotropic fixed point}

We study the stability of the anisotropic inflationary solution 
by considering small perturbations $\delta x_1$, $\delta x_2$, 
and $\delta x_3$ about the anisotropic critical point (ii) 
given by $(x_1^{(c)}, x_2^{(c)}, x_3^{(c)})$, i.e., 
\begin{equation}
x_i=x_i^{(c)}+\delta x_i \qquad (i=1,2,3).
\end{equation}
We expand the function $g(Y)$ around 
$Y_c=(x_1^{(c)}/x_2^{(c)})^2 M_{\rm pl}^4$, i.e., 
\begin{equation}
g(Y)=g_c+g'(Y_c) (Y-Y_c)+\frac{g''(Y_c)}{2} (Y-Y_c)^2
+\cdots\,,
\end{equation}
where $g_c \equiv g(Y_c)$ and $g'(Y)=dg(Y)/dY$. 
Taking the terms up to the second 
order of $Y-Y_c$, we have $\delta P_{,X}=(2g_c'+Y_c g_c'')\delta Y$ 
and $\delta g_1=(g_c'+Y_c g_c'')\delta Y$.
Note that $g_c''$ and $\delta Y$ can be expressed 
as $g_c''=(A^{-1}-g_c-5Yg_c')/(2Y_c^2)$ and 
$\delta Y/M_{\rm pl}^4=2[x_1^{(c)} \delta x_1/(x_2^{(c)})^2
-(x_1^{(c)})^2 \delta x_2/(x_2^{(c)})^3]$.
In the following we omit the subscripts ``$c$'' and 
``$(c)$'' for the background quantities. 

Perturbing Eqs.~(\ref{auto1})-(\ref{auto3}) around the critical point (ii), 
we can write the resulting perturbation equations in the form 
\begin{eqnarray}
\frac{d }{d N}
\left(
\begin{array}{c}
\delta x_1 \\
\delta x_2 \\
\delta x_3
\end{array}
\right) = {\cal M} \left(
\begin{array}{c}
\delta x_1 \\
\delta x_2 \\
\delta x_3
\end{array}
\right) \,,
\label{uvdif}
\end{eqnarray}
where ${\cal M}$ is the $3 \times 3$ matrix expressed 
in terms of $x_1$, $Y$, $A$, $\lambda$, and $\mu$. 
Using the relations (\ref{auani1}), the three eigenvalues of the matrix ${\cal M}$, 
which determine the stability of the anisotropic point (ii), are 
\begin{equation}
\gamma_1=\frac{\sqrt{6}}{2}\lambda x_1-3\,,\qquad 
\gamma_2=\frac{\sqrt{6}}{4}\lambda x_1-\frac32+\frac18 \sqrt{{\cal D}}\,,\qquad
\gamma_3=\frac{\sqrt{6}}{4}\lambda x_1-\frac32-\frac18 \sqrt{{\cal D}}\,,
\label{gam}
\end{equation}
where 
\begin{eqnarray}
{\cal D} =
16 \left[ 9- \sqrt{6}\left(2\lambda+3\mu \right) x_1 \right]^2
+3A (\lambda+2\mu)
\left[ 3 \left( \lambda+2\mu \right) x_1 -2\sqrt{6}\right] 
\left[ \left( \lambda^2 +28\mu\lambda +36\mu^2 \right) x_1 
-2\sqrt{6} \left( \lambda + 14 \mu \right)\right] \,.
\label{deter}
\end{eqnarray}

As long as the condition (\ref{con1}) of the cosmic acceleration is satisfied, 
we have that $\gamma_1<0$. The term $\sqrt{6}\lambda x_1/4-3/2$ 
inside $\gamma_2$ and $\gamma_3$ is also negative under the same condition. 
If ${\cal D}$ is negative, then the anisotropic fixed point is a stable spiral.
For positive ${\cal D}$ the eigenvalue $\gamma_3$ is negative.
When $x_1=2\sqrt{6}/[3(\lambda+2\mu)]$, the eigenvalue 
$\gamma_2$ vanishes for the same signs of $\lambda$ and $\mu$.
In order to see this more precisely, we substitute $x_1=
2\sqrt{6}/[3(\lambda+2\mu)]+\delta$ into the eigenvalue $\gamma_2$, 
where $\delta$ is a small parameter. It then follows that 
\begin{equation}
\gamma_2=-\frac{3\sqrt{6}}{16} (\lambda+2\mu)
\left[ 4+A(\lambda+2\mu)(\lambda+4\mu) \right]\delta
+O(\delta^2)\,.
\end{equation}

Provided that $A>0$, we have $\gamma_2<0$ either for $\lambda>0$, 
$\mu>0$, $\delta>0$ or $\lambda<0$, $\mu<0$, $\delta<0$.
Then the anisotropic fixed point is stable for 
$3(\lambda+2\mu)x_1>2\sqrt{6}$, which is exactly equivalent  
to the condition (\ref{con2}). 
Plugging $x_1=2\sqrt{6}/[3(\lambda+2\mu)]+\delta$ into
$P_{,X}$ of Eq.~(\ref{auani1}), we obtain
\begin{equation}
P_{,X}=\frac14 \lambda (\lambda+2\mu)-\frac{3\sqrt{6}}{32}
(\lambda+2\mu)^3 \delta+O(\delta^2)\,,
\end{equation}
which is positive at $x_1=2\sqrt{6}/[3(\lambda+2\mu)]$ for 
the same signs of $\lambda$ and $\mu$. 
If $P_{,X}>0$ and $A>0$ then the sound speed squared
$c_s^2=P_{,X}A$ is positive, so that the Laplacian instability 
of small-scale perturbations can be avoided.
For $x_1$ away from $2\sqrt{6}/[3(\lambda+2\mu)]$, 
the quantity $P_{,X}$ can be negative.
In order to avoid this, we require
the condition (\ref{con3}).
We also note that $A$ can change its sign at some 
value of $x_1$. Since this depends on the forms of the 
function $g(Y)$, we shall study this property 
in several different models in Sec.~\ref{concretesec}.

We recall that $x_3$ and $x_4^2$ exactly vanish at 
$3(\lambda+2\mu)x_1=2\sqrt{6}$.
In order to keep the small level of anisotropies 
($x_3 \ll 1$ and $x_4^2  \ll 1$), it is required that $x_1$ is 
only slightly larger than the critical value 
$2\sqrt{6}/[3(\lambda+2\mu)]$ for positive 
$\lambda$ and $\mu$.
In this regime the stability of the anisotropic fixed 
point is ensured for $A>0$.


%
\section{Concrete models of power-law inflation}
\label{concretesec}

In this section we study the existence of anisotropic fixed points
as well as their stabilities in concrete models of power-law inflation.
For simplicity we shall focus on the case of the positive 
values of $\lambda$ and $\mu$.

\subsection{Canonical field with an exponential potential}
\begin{figure}
\includegraphics[height=3.4in,width=3.6in]{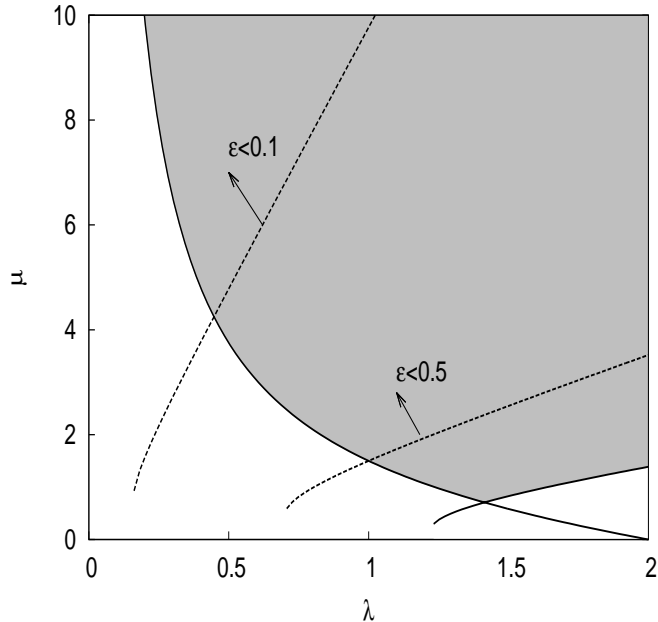}
\caption{\label{fig1}
The parameter space in the $(\lambda,\mu)$ plane for 
the model $P=X-c M_{\rm pl}^4 e^{-\lambda \phi/M_{\rm pl}}$.
The two solid curves, which determine the minimum values of $\mu$ 
for large and small $\lambda$, correspond to the bounds (\ref{canocon1}) 
and (\ref{canocon2}), respectively.
The two dotted curves correspond to $\epsilon=0.1$ and $\epsilon=0.5$. 
In order to realize $\epsilon\ll1$, we require that $\mu/\lambda \gg 1$.}
\end{figure}

Let us first consider the model 
\begin{equation}
P=X-c M_{\rm pl}^4 e^{-\lambda \phi/M_{\rm pl}}
\qquad (c={\rm constant}),
\end{equation}
i.e., the function $g(Y)=1-cM_{\rm pl}^4/Y$.
Solving the first two equations of (\ref{auani1}) for this function, 
we obtain the following anisotropic fixed point 
\begin{eqnarray}
& & x_1=\frac{2\sqrt{6}(\lambda+2\mu)}
{\lambda^2+8\mu \lambda+12\mu^2+8}\,,\qquad 
c x_2^2=\frac{6(2+2\mu^2+\mu \lambda)
(8+12\mu^2+4\mu \lambda-\lambda^2)}
{(\lambda^2+8\mu \lambda+12\mu^2+8)^2}\,,\nonumber \\
& & x_3=\frac{2(\lambda^2+2\mu \lambda-4)}
{\lambda^2+8\mu \lambda+12\mu^2+8}\,,\qquad
x_4^2=\frac{3(\lambda^2+2\mu \lambda-4)
(8+12\mu^2+4\mu \lambda-\lambda^2)}
{(\lambda^2+8\mu \lambda+12\mu^2+8)^2}\,,
\label{xiexp}
\end{eqnarray}
which agree with those derived in Ref.~\cite{Kanno10}. 
The upper bound of Eq.~(\ref{con1}) translates to 
\begin{equation}
8+12\mu^2-4\mu \lambda-5\lambda^2>0\,,
\label{canocon1}
\end{equation}
which is satisfied for $\mu \gg \lambda$. 
The condition (\ref{con2}) for the existence of the anisotropic 
fixed point is interpreted as 
\begin{equation}
\lambda^2+2\mu \lambda-4>0\,.
\label{canocon2}
\end{equation}
Since $P_{,X}=1>0$, $x_1$ is smaller 
than the upper bound of Eq.~(\ref{x1confin}). 
In this model the quantity $A$ is 1, so that the stability 
of the anisotropic inflationary solution is ensured under 
the condition (\ref{canocon2}) in the regime where
$x_1$ is not far away from the value $2\sqrt{6}/[3(\lambda+2\mu)]$. 
Even for $x_1 \gg 2\sqrt{6}/[3(\lambda+2\mu)]$ the determinant 
${\cal D}$ appearing in $\gamma_2$ of Eq.~(\ref{gam}) becomes 
negative and hence the fixed point is a stable spiral.
This means that the anisotropic inflationary solution is an attractor 
under the condition (\ref{canocon2}) \cite{Kanno10}.

In Fig.~\ref{fig1} we show the viable parameter space in the 
$(\lambda,\mu)$ plane satisfying the two bounds 
(\ref{canocon1}) and (\ref{canocon2}).
The stable anisotropic inflation can be realized for the parameters 
in the shaded region.
We also plot the two curves corresponding to $\epsilon=0.1$ 
and $\epsilon=0.5$.
For $\lambda$ and $\mu$ satisfying the conditions $\mu \gg \lambda$ 
and $\mu \gg 1$, we approximately have $x_1 \simeq \sqrt{6}/(3\mu)$ from 
Eq.~(\ref{xiexp}) and hence $\epsilon \simeq \lambda/\mu$ from Eq.~(\ref{dotH3}).
The slow-roll parameter $\epsilon$ of the order of $10^{-2}$ can be 
realized for $\mu/\lambda=O(10^2)$. 
If $\mu/\lambda=10^2$, for example, the condition 
$\lambda^2+2\mu \lambda-4>0$
translates to $\mu=10^2 \lambda>14$.

In the limit that $\lambda \to 0$, the condition $\lambda^2+2\mu \lambda-4>0$ 
is not fulfilled. Hence, in this model, the stable anisotropic solution 
does not exist on the de Sitter background. 
This comes from the fact that  the field is frozen 
in the slow-roll limit ($\epsilon \to 0$), 
so that there is no variation of the coupling $f(\phi)$ 
in Eq.~(\ref{fphi}) to give rise to anisotropic solutions.

\subsection{Generalized ghost condensate}

The second model is the generalized ghost condensate given by 
the Lagrangian 
\begin{equation}
P=-X+\frac{c}{M_{\rm pl}^{4n}}e^{n\lambda \phi/M_{\rm pl}}X^{n+1}
\qquad (c,n={\rm constant~with}~n \geq 1),
\end{equation}
in which case $g(Y)=-1+c(Y/M_{\rm pl}^4)^n$.
The diatonic ghost condensate model \cite{Piazza} corresponds to the case
$n=1$. {}From the first two equations of (\ref{auani1}) we find that 
\begin{equation}
x_1=\frac{-\sqrt{6}[3\lambda+2\mu+(5\lambda+6\mu)n]\pm 
\sqrt{6[3\lambda+2\mu+(5\lambda+6\mu)n]^2+48(n+1)
[2(4-\lambda^2-5\lambda\mu-6\mu^2)n-\lambda(\lambda+2\mu)]}}
{2[2(4-\lambda^2-5\lambda\mu-6\mu^2)n-\lambda(\lambda+2\mu)]} \,.
\label{x1so}
\end{equation}
Since the plus sign of Eq.~(\ref{x1so}) can give positive values of $x_1$, 
we use this solution in the following discussion.
Then the anisotropic parameter $x_3$ reads
\begin{equation}
x_3=\frac{3\lambda+10\mu+(\lambda+6\mu)n-\sqrt{(9\lambda^2
-20\lambda\mu-60\mu^2+64)n^2
+2(3\lambda^2-20\lambda\mu-36\mu^2+32)n+(\lambda-2\mu)^2}}
{3\lambda+2\mu+(5\lambda+6\mu)n+\sqrt{(9\lambda^2-20\lambda\mu
-60\mu^2+64)n^2
+2(3\lambda^2-20\lambda\mu-36\mu^2+32)n+(\lambda-2\mu)^2}}\,.
\label{x3so}
\end{equation}

The condition (\ref{x1confin}) translates to
\begin{equation}
\mu_1<\mu<\mu_2 \,,\quad {\rm where} \quad
\mu_1 \equiv \frac{\sqrt{(2n+1)^2\lambda^2
+24n(n+1)}-(n+2)\lambda}{6(n+1)}\,,\quad 
\mu_2 \equiv \frac{1}{12} \left( \lambda+\sqrt{\lambda^2+
\frac{96n}{n+1}}\right)\,.
\label{dilacon1}
\end{equation}
{}From the condition (\ref{con1}) the variable $\mu$ is bounded to be
\begin{equation}
\mu<\mu_3\,,\quad {\rm where} \quad 
\mu_3 \equiv
\frac{(2n+1)\lambda+\sqrt{24n^2-(11n^2+26n-1)\lambda^2}}{6n}\,.
\label{dilacon2}
\end{equation}
For the determinant of Eq.~(\ref{x1so}) to be positive, 
we require that   
\begin{equation}
\mu<\mu_4\,,\quad {\rm where} \quad
\mu_4 \equiv
\frac{4\sqrt{2n(5n+1)(n+1)^2
\lambda^2+4n(n+1)(15n^2+18n-1)}-(5n^2+10n+1)\lambda}
{2(15n^2+18n-1)}\,.
\label{dilacon3}
\end{equation}
\begin{figure}
\includegraphics[height=3.3in,width=3.5in]{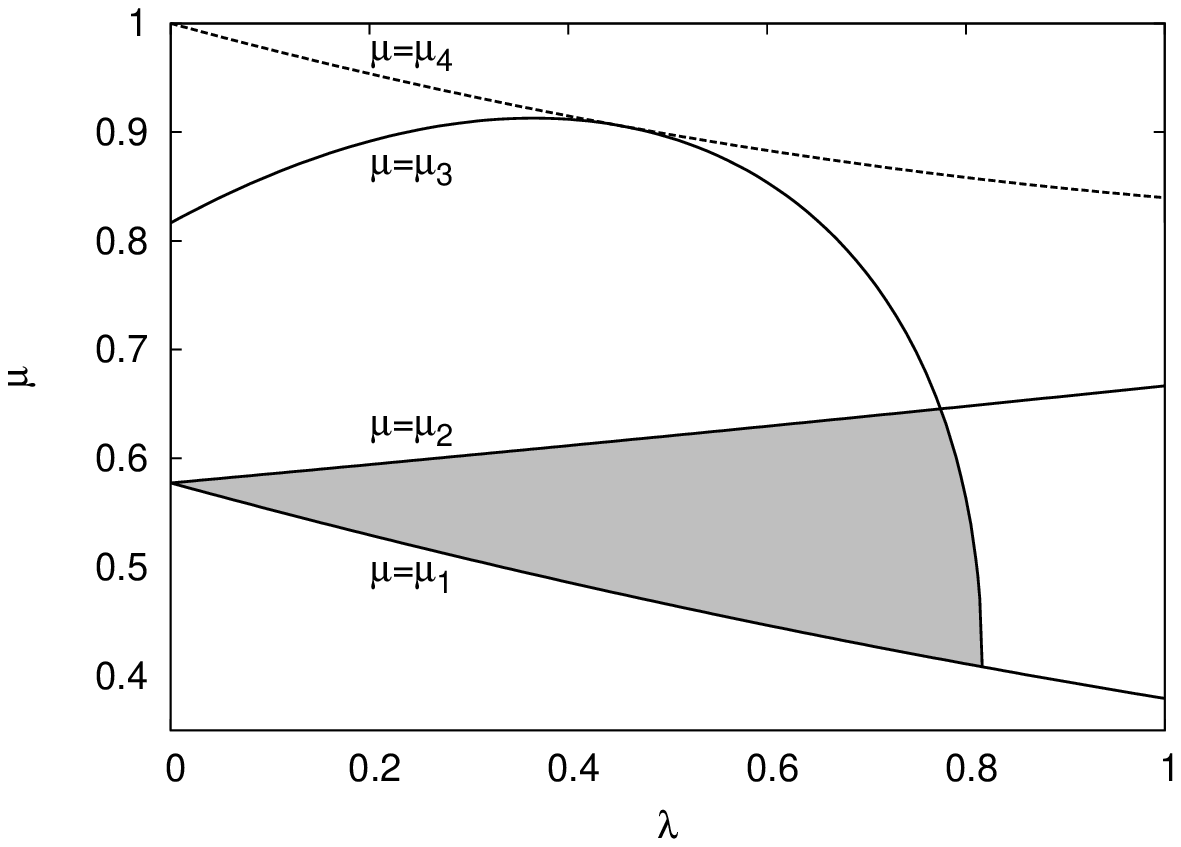}
\includegraphics[height=3.3in,width=3.5in]{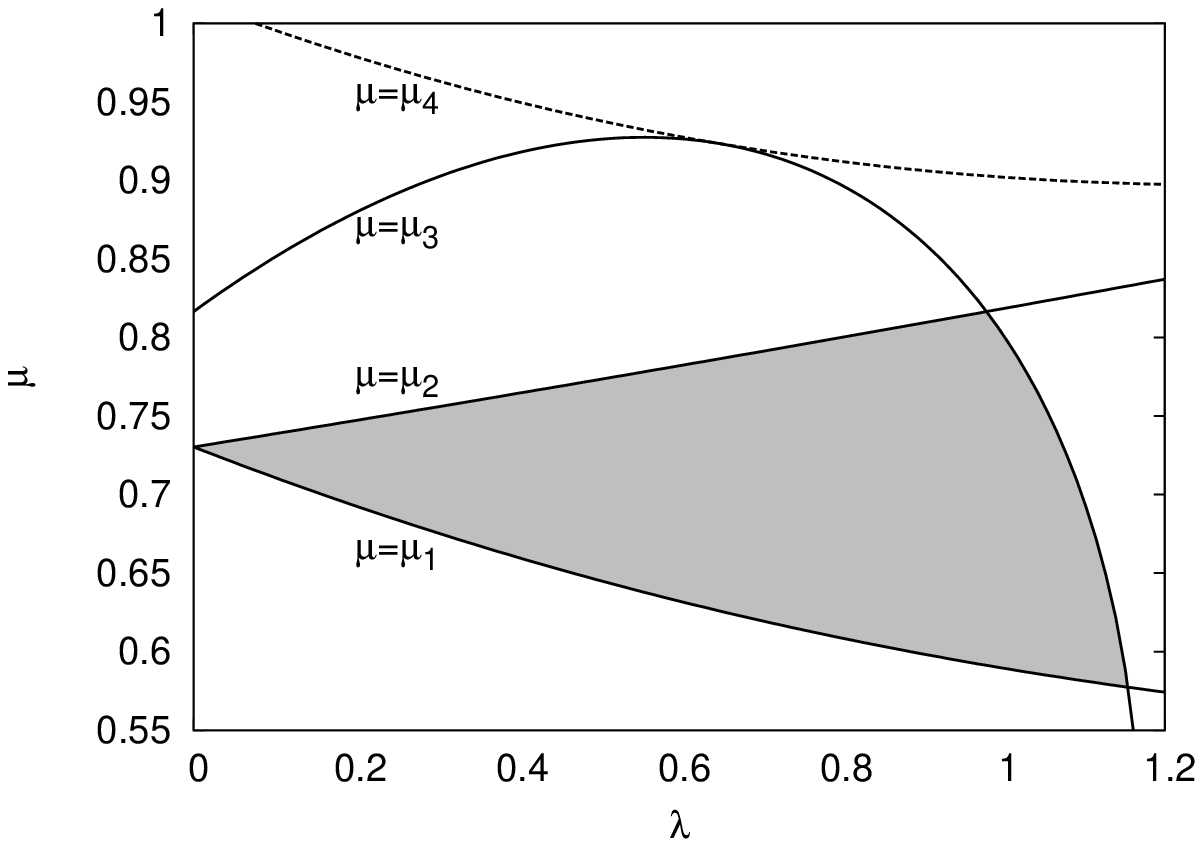}
\caption{\label{figab}
The parameter space in the $(\lambda,\mu)$ plane for the generalized 
ghost condensate model with $n=1$ (left) and $n=4$ (right).
The four curves correspond to the borders given 
in Eqs.~(\ref{dilacon1}), 
(\ref{dilacon2}), and (\ref{dilacon3}). In the shaded region, 
all the conditions (\ref{dilacon1})-(\ref{dilacon3}) are satisfied.}
\end{figure}

In Fig.~\ref{figab} we plot the parameter space in the 
$(\lambda,\mu)$ plane satisfying the conditions 
(\ref{dilacon1})-(\ref{dilacon3}) for $n=1$ and $n=4$.
In the limit that $\lambda \to 0$, the region described 
by (\ref{dilacon1}) shrinks to the point 
$\mu=\sqrt{2n/[3(n+1)]}$.  
As we see in Fig.~\ref{figab}, the region (\ref{dilacon1})
tends to be wider for larger $\lambda$. 
The condition (\ref{dilacon2}) gives upper bounds 
of $\lambda$ and $\mu$. 
The intersection point of the curves $\mu=\mu_1$ 
and $\mu=\mu_3$ is given by 
$(\lambda,\mu)=(\sqrt{2n/(n+2)},\sqrt{n/[2(n+2)]})$,  
whereas the curves $\mu=\mu_2$ and 
$\mu=\mu_3$ intersect at the point 
$(\lambda,\mu)=(\sqrt{6n/[5(n+1)]},\sqrt{5n/[6(n+1)]})$.
For $n \geq 1$ the parameters $\lambda$ and $\mu$ 
are in the range
\begin{equation}
0<\lambda<\sqrt{\frac{2n}{n+2}}\,,\qquad
\sqrt{\frac{n}{2(n+2)}}<\mu<\sqrt{\frac{5n}{6(n+1)}}\,.
\label{lammucon}
\end{equation}
We note that the condition (\ref{dilacon3}) does 
not provide an additional bound. 
{}From Eq.~(\ref{lammucon}) the parameter $\mu$ 
is of the order of $0.1$ (with the maximum value
$\mu=\sqrt{5/6}$ in the limit $n \to \infty$). 

\begin{figure}
\includegraphics[height=3.4in,width=3.6in]{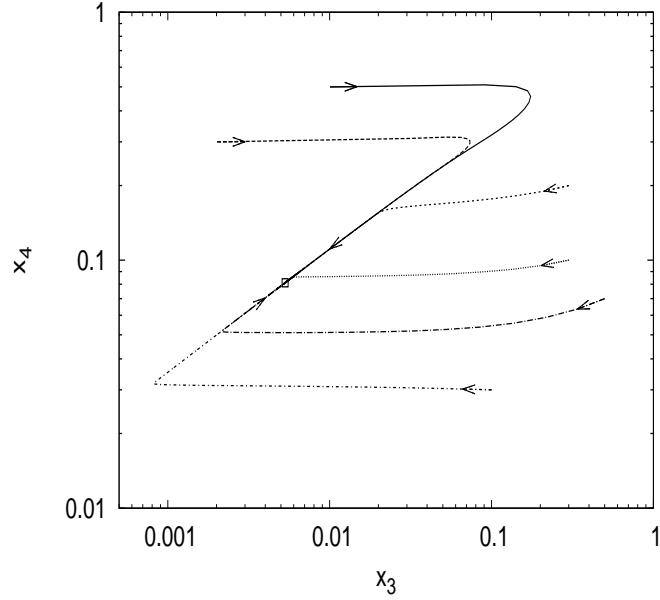}
\caption{\label{fig3}
The phase space in the two-dimensional plane ($x_3,x_4$) for 
the dilatonic ghost condensate model with the Lagrangian 
$P=-X+e^{\lambda \phi/M_{\rm pl}}X^2/M_{\rm pl}^4$. 
The model parameters are chosen to be $\lambda=0.35$ and 
$\mu=0.5$ with the initial condition $x_1=1.0$ and 
several different initial values of $x_2$ and $x_3$.
The solutions finally converge to the anisotropic fixed point
$(x_3,x_4)=(5.306 \times 10^{-3},8.109 \times 10^{-2})$ with 
$x_1=1.216$, $x_2=1.629$, and $\epsilon=0.521$.}
\end{figure}

In Fig.~\ref{fig3} we plot the phase space trajectories in the 
two-dimensional plane ($x_3,x_4$) for $n=1$, 
$\lambda=0.35$, and $\mu=0.5$. 
The trajectories with different initial conditions converge 
to the anisotropic fixed point (ii) and 
hence the fixed point is stable.
As long as $\mu$ is close to the lower bound $\mu=\mu_1$,  
the anisotropic parameter $x_3=\Sigma/H$ is 
much smaller than 1.
For increasing $\mu$ the anisotropy gets larger.
In the numerical simulation of Fig.~\ref{fig3} the slow-roll parameter is 
$\epsilon=0.521$ along the anisotropic attractor.
In order to realize $\epsilon$ of the order of $10^{-2}$, we require 
that $\lambda=O(10^{-2})$.

For $\mu$ close to its upper bound, it can happen 
that the stability of the anisotropic fixed point is subject to change.
In fact, the parameter $A=1/[c(n+1)(2n+1)(Y/M_{\rm pl}^{4})^n-1]$ 
diverges at $c(Y/M_{\rm pl}^{4})^n=1/[(n+1)(2n+1)]$.
This leads to the sign change of the determinant 
(\ref{deter}) from negative to positive by passing the 
singular point at $\mu=\mu_5$.
If $n=1$ then we have $\mu_5=\sqrt{\lambda^2+8}/3-\lambda/6$, 
so the anisotropic fixed point is stable for 
\begin{equation}
\mu<\sqrt{\lambda^2+8}/3-\lambda/6\,.
\label{mu5}
\end{equation}
This does not give an additional bound to those given in 
Eqs.~(\ref{dilacon1})-(\ref{dilacon3}).
When $n>1$ the condition (\ref{mu5}) is more involved, but  
the situation is similar to that discussed for $n=1$. 
It is worth mentioning that, for $n=1$, the condition (\ref{mu5}) 
is equivalent to $x_3<1$.

The maximum value of $x_3$ is reached for 
$(\lambda,\mu)=(\sqrt{6n/[5(n+1)]},\sqrt{5n/[6(n+1)]})$.
Substituting these values into Eq.~(\ref{x3so}) we have
$x_3=1/3$, which corresponds to the upper bound of 
(\ref{x3upper}) with $\mu/\lambda=5/6$. 
Hence the anisotropic parameter is constrained to be 
\begin{equation}
\Sigma/H<1/3\,,
\label{aniupper}
\end{equation}
which holds independent of $n$. 
This bound comes from the combination of the conditions $P_{,X}>0$ 
and $\lambda x_1<\sqrt{6}/3$. 
If we impose the NEC $\rho_t+P_t>0$ instead of $P_{,X}>0$, 
the upper bound (\ref{aniupper}) gets larger.
However, such a large anisotropy is not accepted observationally. 

In summary, for $n \geq 1$, there exist the allowed parameter spaces 
satisfying all the conditions (\ref{dilacon1})-(\ref{dilacon3}).
In order to realize the sufficient amount of inflation ($\epsilon \ll 1$) 
with the suppressed anisotropy ($x_3 \ll 1$), we require that 
$\lambda \ll 1$ and that $\mu$ is close to 
the lower bound $\mu_1$.

\subsection{DBI model}

The DBI model is characterized by the Lagrangian \cite{DBI}
\begin{equation}
P=-h(\phi)^{-1} \sqrt{1-2h(\phi)X}
+h(\phi)^{-1}-V(\phi)\,,
\label{DBIlag}
\end{equation}
where $h(\phi)$ and $V(\phi)$ are functions of $\phi$.
For the choice $g(Y)=-(m^4/Y)\sqrt{1-2Y/m^4}-M^4/Y$, where 
$m$ and $M$ are constants having a dimension of mass, 
we obtain the Lagrangian (\ref{DBIlag}) with $h(\phi)^{-1} =
m^4e^{-\lambda \phi/M_{\rm pl}}$ and 
$V(\phi)=(M^4+m^4)e^{-\lambda \phi/M_{\rm pl}}$.
The ultra-relativistic regime corresponds to the case 
where the quantity $Y/m^4$ is close to $1/2$.
In order to sustain sufficient amount of inflation 
in this regime, we require that the ratio 
$c_M \equiv M^4/m^4$ is much larger than 1 \cite{Ohashi2011}.

Since $P_{,X}=[1-2h(\phi)X]^{-1/2}>0$ in the DBI model, the upper 
bound of Eq.~(\ref{x1confin}) and the NEC are automatically satisfied. 
We also have $A=(1-2Y/m^4)^{3/2}$, so that there is no divergence 
associated with the determinant (\ref{deter}) in the regime $Y/m^4<1/2$.
{}From the first two equations of (\ref{auani1}) we find that 
the anisotropic fixed point satisfies the fourth order equation 
of $x_1$, but it is not analytically solvable 
for general values of $\lambda$, $\mu$, and $c_M$. 
However, substituting the lower bound of Eq.~(\ref{x1confin}) 
into the fourth order equation of $x_1$, 
we obtain the following constraint 
\begin{equation}
\mu>\frac{2\sqrt{\lambda^4+12c_M\lambda^2+36}
-\lambda^2}{6\lambda}\,.
\label{DBImu}
\end{equation}
In the ultra-relativistic regime the quantity $Y/m^4$ is close to $1/2$, 
so that $P_{,X}=(1-2Y/m^4)^{-1/2}$ is much larger than 1. 
Using the bound (\ref{con2}), the anisotropic fixed point 
of Eq.~(\ref{auani1}) satisfies the relation
$P_{,X}<\lambda (\lambda+2\mu)/4$, i.e., 
\begin{equation}
\sqrt{1-\frac{2Y}{m^4}}>\frac{4}{\lambda (\lambda+2\mu)}\,.
\label{Ycon}
\end{equation}

In order to realize the situation where $Y/m^4$ is close to $1/2$,
we require that $\lambda (\lambda+2\mu) \gg 1$. 
As $x_1$ is away from the value $2\sqrt{6}/[3(\lambda+2\mu)]$, 
there is a tendency that the anisotropic fixed point 
deviates from the ultra-relativistic regime because of 
the decrease of $P_{,X}$. 
In the following we focus on the situation 
where $x_1$ is close to $2\sqrt{6}/[3(\lambda+2\mu)]$, 
in which case the anisotropic fixed point is stable with 
a small anisotropy.

For $x_1 \simeq 2\sqrt{6}/[3(\lambda+2\mu)]$, the slow-roll parameter 
is given by $\epsilon \simeq 2\lambda/(\lambda+2\mu)$ from Eq.~(\ref{dotH3}).
In order to realize $\epsilon \ll 1$, we need the condition 
$\mu \gg \lambda$. 
Then, the condition $\lambda (\lambda+2\mu) \gg 1$
discussed above can be interpreted as $\mu \lambda \gg 1$.
{}From Eq.~(\ref{DBImu}) the condition $\mu \lambda \gg 1$ 
can be satisfied for $c_M\lambda^2 \gg 10$, in which case 
Eq.~(\ref{DBImu}) reduces to $\mu>2\sqrt{3c_M}/3$.
When $c_M={\cal O}(100)$, for example, we have 
$\mu  \gtrsim {\cal O}(10)$ and $\lambda \gtrsim {\cal O}(1)$. 

\begin{figure}
\includegraphics[height=4.0in,width=4.2in]{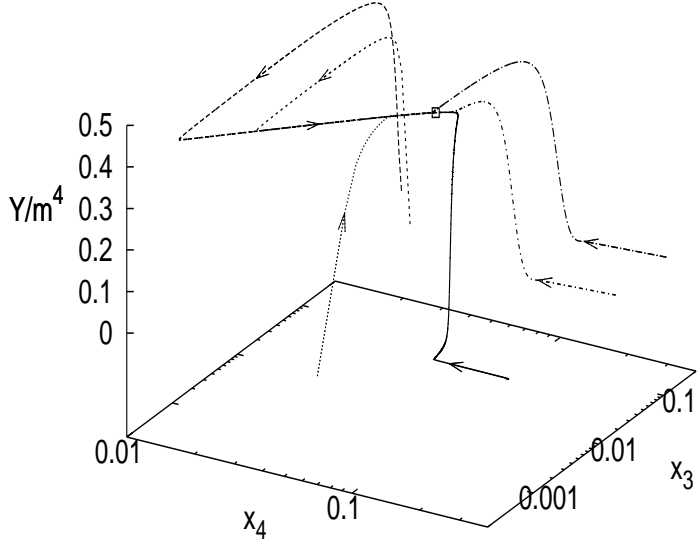}
\caption{\label{fig4}
The three-dimensional phase space ($Y/m^4,x_3,x_4$) for 
the DBI model with the Lagrangian 
$P=-m^4 e^{-\lambda \phi/M_{\rm pl}}\sqrt{1-2Xe^{\lambda \phi/M_{\rm pl}}/m^4}
-M^4 e^{-\lambda \phi/M_{\rm pl}}$. 
The model parameters are chosen to be 
$c_M=M^4/m^4=500$, $\lambda=1$, and $\mu=26$
with the initial condition $Y/m^4=10^{-2}$ 
and several different initial values of $x_3$ and $x_4$.
The trajectories with different initial conditions converge 
to the anisotropic fixed point 
$(Y/m^4,x_3,x_4)=
(4.911 \times 10^{-1}, 5.494 \times 10^{-3},9.020 \times 10^{-2})$ with 
$x_1=3.098 \times 10^{-2}$ and $\epsilon=3.794 \times 10^{-2}$.}
\end{figure}

For compatibility of the two conditions (\ref{con1}) 
and (\ref{con2}), we require that $\mu>\lambda/2$. 
If $\lambda>\lambda_m \equiv \sqrt{2c_M+2\sqrt{c_M^2+3}}$, 
the condition $\mu>\lambda/2$ is stronger than the bound (\ref{DBImu}). 
In the regime $\lambda<\lambda_m$, as $\lambda$ gets larger 
around the lower bound of $\mu$ given in Eq.~(\ref{DBImu}), 
the slow-roll parameter also increases and 
it reaches the value $\epsilon=1$ at $\lambda=\lambda_m$.
Then, the realization of anisotropic inflation
demands the condition 
\begin{equation}
\lambda<\sqrt{2c_M+2\sqrt{c_M^2+3}}\,.
\label{lamcon}
\end{equation}
When $c_M=500$, for example, the condition (\ref{lamcon}) 
translates to $\lambda<44.7$.
As long as $\lambda$ is much smaller than the upper bound of 
Eq.~(\ref{lamcon}), anisotropic inflation with $\epsilon \ll 1$
occurs in the ultra-relativistic regime for $\mu$ close 
to the lower bound of Eq.~(\ref{DBImu}).

In Fig.~\ref{fig4} we show the trajectories of solutions in the 
three-dimensional phase space $(Y/m^4, x_3,x_4)$ 
for $c_M=500$, $\lambda=1$, and $\mu=26$. 
In this case the solutions with several different initial conditions 
converge to the anisotropic fixed point with constant values 
of $x_3$, $x_4$ satisfying $x_3 \ll 1$ and $x_4 \ll 1$.
The attractor is in the ultra-relativistic regime ($Y/m^4$ close to $1/2$)
with $\epsilon$ of the order of 0.01.
It is also possible to realize stable anisotropic inflation for 
$\lambda=O(10)$ and $\mu=O(10)$, but in such cases the slow-roll 
parameter $\epsilon$ is not much smaller than 1.

In summary, the stable anisotropic DBI inflation can be realized 
in the ultra-relativistic regime under the conditions 
(\ref{DBImu}) and (\ref{lamcon}) for $\mu$ close to 
the lower bound (\ref{DBImu}).


\section{Conclusions}
\label{consec}

We have studied the dynamics of anisotropic power-law
k-inflation in the presence of a vector kinetic term 
$F_{\mu \nu}F^{\mu \nu}$ coupled to the inflaton field $\phi$.
Such a power-law k-inflation can be accommodated for the general 
Lagrangian $P=Xg(Y)$, where $Y=Xe^{\lambda \phi/M_{\rm pl}}$. 
The cosmological dynamics in the anisotropic cosmological background
is known by solving the autonomous equations 
(\ref{auto1})-(\ref{auto3}).

Without specifying the functional forms of $g(Y)$, we have shown 
that anisotropic inflationary solutions exist for the 
exponential coupling (\ref{fphi}).
The anisotropic fixed point satisfying Eq.~(\ref{auani1}) is 
present for $3(\lambda+2\mu)x_1>2\sqrt{6}$, where 
$x_1=\dot{\phi}/(\sqrt{6}HM_{\rm pl})$. 
The condition for the cosmic acceleration 
translates to $\lambda x_1<\sqrt{6}/3$. 
Provided the conditions $3(\lambda+2\mu)x_1>2\sqrt{6}$ and 
$A=(P_{,X}+2XP_{,XX})^{-1}>0$ are satisfied, 
the anisotropic inflationary fixed point is stable 
in the regime where $x_1$ is close to 
$2\sqrt{6}/[3(\lambda+2\mu)]$. 
This property holds irrespective of the forms of $g(Y)$ and 
hence the anisotropic hair survives whenever the anisotropic 
power-law inflationary solutions are present. 

The quantity $A$ is related to the sound speed $c_s$ as 
$c_s^2=AP_{,X}$, so that the Laplacian instability can 
be avoided for $A>0$ and $P_{,X}>0$.
For the models in which $P_{,X}$ can be negative, 
it happens that the NEC $\rho_t+P_t>0$ is not satisfied 
for some model parameters, see Eq.~(\ref{nec}). 
In the de Sitter limit ($\lambda \to 0$) 
we found that the NEC is always violated 
for anisotropic solutions. This is consistent 
with the Wald's cosmic no hair conjecture. 
As long as $\lambda$ is not 0, there are some 
parameter spaces in which the NEC is satisfied.

In Sec.~\ref{concretesec} we applied our general results to 
concrete models of k-inflation such as the generalized ghost 
condensate and the DBI model. 
In the generalized ghost condensate 
we showed that there are allowed parameter spaces in the 
$(\lambda,\mu)$ plane where stable anisotropic
inflationary solutions with $P_{,X}>0$ and $A>0$ 
are present, see Fig.~\ref{figab}. The existence of such 
anisotropic attractors is confirmed in the numerical 
simulation of Fig.~\ref{fig3}. 
In this model anisotropic inflation with $\lambda=O(0.1)$ 
and $\mu=O(0.1)$ occurs, but if the slow-roll parameter 
$\epsilon$ is of the order of $10^{-2}$, 
it follows that $\lambda=O(10^{-2})$. 
In the DBI model there exists stable anisotropic inflationary solutions 
in the ultra-relativistic regime ($Y/m^4 \simeq 1/2$) 
for $\mu$ close to the lower bound of Eq.~(\ref{DBImu}) and 
$\lambda$ satisfying the bound (\ref{lamcon})
(see Fig.~\ref{fig4}).
The model parameters are typically of the order of 
$\lambda=O(1)$ and $\mu=O(10)$
to realize $\epsilon=O(10^{-2})$.

While we focused on the vector field coupled to the inflaton 
in this paper, we expect that the similar property 
should also hold for the two-form field models
studied in Ref.~\cite{Ohashi} in the context 
of potential-driven slow-roll inflation. 
It is also known that in k-inflation the non-Gaussianities of 
scalar metric perturbations can be large for the equilateral 
shape due to the non-linear field self-interactions 
inside the Hubble radius \cite{nongau}. 
It will be of interest to study how the non-linear 
estimator $f_{\rm NL}$ of the single-field k-inflation 
is modified by the interactions between 
inflaton and the vector/two-form fields. 
We leave these issues for future work.

\acknowledgements

This work is supported by the Grant-in-Aid for Scientific Research 
Fund of the Ministry of Education, Science and Culture of Japan 
(Nos.~23$\cdot$6781, 25400251, and 24540286), the Grant-in-Aid 
for Scientific Research on Innovative Area (No.~21111006).

\section*{Appendix}
\label{appe}
 
In this Appendix we provide more explicit analysis for the properties 
of isotropic and anisotropic solutions given in 
Eqs.~(\ref{auiso}) and (\ref{auani1}).

The power-law inflationary solution corresponds to $\dot{\alpha}=H=\zeta/t$, 
where $\zeta$ is a constant larger than 1. 
Since the quantities $x_3=\dot{\sigma}/H$ and $x_1=\dot{\phi}/(\sqrt{6}HM_{\rm pl})$ 
are constant along the fixed points, we have that 
$\dot{\sigma}=\eta/t$ and $\dot{\phi}/M_{\rm pl}=\xi/t$, respectively, 
where $\eta=\zeta x_3$ and $\xi=\sqrt{6}x_1 \zeta$.
Then, the evolution of $\alpha$, $\sigma$, and $\phi$ 
is characterized by 
\begin{equation}
\alpha = \zeta \log \frac{t}{t_0}\,,\qquad 
\sigma = \eta \log \frac{t}{t_0}\,, \qquad 
\frac{\phi}{M_{\rm pl}}  = \xi \log \frac{t}{t_0}\,,
\label{poweras}
\end{equation}
where $t_0$ is a constant.
For $Y=Xe^{\lambda \phi/M_{\rm pl}}$ to be constant, 
we need to require
\begin{equation}
\lambda \xi =2\,.
\label{scaling}
\end{equation}
For the solutions (\ref{poweras}) satisfying the relation (\ref{scaling}), 
the dimensionless variables defined in Eq.~(\ref{dimendef}) read
\begin{equation}
x_1=\frac{2}{\sqrt{6}\lambda \zeta}\,,\qquad
x_2=\frac{M_{\rm pl}t_0}{\sqrt{3}\zeta}\,,\qquad
x_3=\frac{\eta}{\zeta}\,,\qquad
x_4^2=\frac{W^2}{6\zeta^2}\,,
\label{xire}
\end{equation}
where $W^2=p_A^2 t_0^2/(M_{\rm pl}^2 f_0^2)$. 
Substituting the solutions (\ref{poweras}) 
into Eqs.~(\ref{be1})-(\ref{be4}), we obtain
\begin{eqnarray}
&& \mu \xi - 2\zeta - 2\eta = -1\,,  \label{similar}\\
&& \zeta^2 = \eta^2 + \frac{2P_{,X}-g}{6} \xi^2 + \frac{W^2}{6}\,, \label{hamilton} \\
&& \zeta = 3\eta^2 + \frac{P_{,X}}{2}\xi^2 + \frac{W^2}{3}\,,  \label{evolution}\\
&& \eta = 3\zeta \eta - \frac{W^2}{3}\,,  \label{aniso}\\
&&  \xi -3AP_{,X}\zeta \xi -\frac{\lambda}{2}\left( 1-A P_X \right) \xi^2 
+ \frac{\lambda}{2} (P_{,X}-g) A\xi^2 -\mu A W^2 = 0 \label{KG}\,.
\end{eqnarray}
Notice that Eq.~(\ref{similar}) follows from the demand to have the 
time dependence $t^{-2}$ for the last term of Eq.~(\ref{be1}).
Plugging the relation (\ref{scaling}) into Eq.~(\ref{KG}), 
it follows that 
\begin{equation}
W^2=\frac{2}{\lambda \mu} \left[ P_{,X} (2-3\zeta)-g \right]\,.
\label{W2simple}
\end{equation}

First, let us seek isotropic solutions. 
In this case, Eq.~(\ref{similar}) is absent and $\eta = W=0$.
{}From Eqs.~(\ref{evolution}) and (\ref{hamilton}) we obtain
the following relations
\begin{equation}
\zeta = \frac{2P_{,X}}{\lambda^2}\,,\qquad
P_{,X}^2 -\frac{\lambda^2}{3} P_{,X} + 
\frac{\lambda^2}{6} g=0 \label{Ysol}\,,
\end{equation}
respectively. Note that these are consistent with Eq.~(\ref{W2simple}). 
On using the correspondence (\ref{xire}), 
we find that the two relations (\ref{Ysol}) are equivalent 
to the first two of Eq.~(\ref{auiso}).

Now, we move on to anisotropic power-law solutions. 
{}From Eq.~(\ref{similar}) we have $\zeta + \eta = 1/2 + \mu/\lambda$.
Combining Eqs.~(\ref{evolution}) and (\ref{aniso}), it follows that 
$\zeta + \eta = 3\eta (\zeta +\eta) + P_{,X} \xi^2/2$. 
Then we obtain
\begin{eqnarray}
\zeta &=& \frac{(\lambda + 2\mu)(\lambda + 6\mu)+ 8 P_{,X}}
{6\lambda (\lambda + 2\mu)}\,,
\label{sol-zeta}\\
\eta &=& \frac{\lambda^2 + 2\lambda \mu -4 P_{,X}}{3\lambda (\lambda +2\mu)}\,,  
\label{sol-eta}
\end{eqnarray}
by which the anisotropy of the expansion is 
\begin{equation}
\frac{\Sigma}{H} = \frac{\eta}{\zeta}
= \frac{2(\lambda^2 +2\lambda \mu -4 P_{,X})}
{(\lambda + 2\mu)(\lambda + 6\mu)+ 8 P_{,X}}\,.
\label{rafi}
\end{equation}
Substituting Eqs.~(\ref{sol-zeta}) and (\ref{sol-eta}) 
into Eq.~(\ref{aniso}), we have
\begin{equation}
W^2 =-\frac{(\lambda^2+2\lambda \mu-4P_{,X})
(\lambda^2-4\lambda \mu-12\mu^2-8P_{,X})}
{2\lambda^2 (\lambda+2\mu)^2}\,.
\label{Wdfi}
\end{equation}

{}From Eq.~(\ref{sol-zeta}) and the first of Eq.~(\ref{xire}) 
we can express $P_{,X}$ in terms of $x_1$. 
This exactly corresponds to the first relation of Eq.~(\ref{auani1}). 
Substituting this into Eqs.~(\ref{rafi})-(\ref{Wdfi}) and 
using the correspondence (\ref{xire}), we obtain 
the third and fourth relations of Eq.~(\ref{auani1}). 
On using Eqs.~(\ref{W2simple}) and (\ref{Wdfi}) as well as 
the relation $P_{,X}=g+g_1$, we find that $g_1$ can be expressed
as the second of Eq.~(\ref{auani1}). 
If we want to obtain the metric explicitly, 
we need to use Eq.~(\ref{xire})
to get the following relations
\begin{equation}
\zeta = \frac{2}{\sqrt{6} \lambda x_1 } \ , 
\qquad \eta = \frac{2 x_3 }{\sqrt{6} \lambda x_1 } \ .
\end{equation}
The anisotropic power-law inflationary solutions are given by
\begin{equation}
ds^2 = - dt^2 + \left(\frac{t}{t_0}\right)^{2\zeta} 
\left[ \left(\frac{t}{t_0}\right)^{-4\eta}dx^2
+\left(\frac{t}{t_0}\right)^{2\eta}(dy^2+dz^2) \right] \ .
\label{anisotropic-metric2}
\end{equation}
Now it is easy to write down the metric corresponding to 
the solutions derived in Sec.~\ref{concretesec}.


\begin{thebibliography}{99}

\bibitem{oldinf} 
A.~A.~Starobinsky, 
Phys.\ Lett.\ B \textbf{91}, 99 (1980);
K.~Sato, Mon.\ Not.\ R.\ Astron.\ Soc. \textbf{195}, 467 (1981); \\
K.~Sato, Phys.\ Lett.\ \textbf{99B}, 66 (1981);\\
D.~Kazanas, 
Astrophys.\ J.\ \textbf{241} L59 (1980); \\
A.~H.~Guth, 
Phys.\ Rev.\ D \textbf{23}, 347 (1981).

\bibitem{oldper} 
V.~F.~Mukhanov and G.~V.~Chibisov, 
JETP Lett.\ \textbf{33}, 532 (1981); \\
A.~H.~Guth and S.~Y.~Pi,
Phys.\ Rev.\ Lett.\ \textbf{49} (1982) 1110; \\
S.~W.~Hawking, Phys.\ Lett.\ B \textbf{115}, 295 (1982); \\
A.~A.~Starobinsky, 
Phys.\ Lett.\ B \textbf{117} (1982) 175; \\
J.~M.~Bardeen, P.~J.~Steinhardt and M.~S.~Turner, 
Phys.\ Rev.\ D \textbf{28}, 679 (1983).

\bibitem{WMAP} 
D.~N.~Spergel {\it et al.}  [WMAP Collaboration],
Astrophys.\ J.\ Suppl.\  {\bf 148}, 175 (2003)
[astro-ph/0302209];\\
G.~Hinshaw {\it et al.}  [WMAP Collaboration],
Astrophys.\ J.\ Suppl.\  {\bf 208}, 19 (2013)
 [arXiv:1212.5226 [astro-ph.CO]].

\bibitem{Planck} 
P.~A.~R.~Ade {\it et al.}  [Planck Collaboration],
arXiv:1303.5076 [astro-ph.CO];\\
P.~A.~R.~Ade {\it et al.}  [Planck Collaboration],
arXiv:1303.5082 [astro-ph.CO].

\bibitem{aniobser} 
N.~E.~Groeneboom and H.~K.~Eriksen,
Astrophys.\ J.\  {\bf 690}, 1807 (2009)
[arXiv:0807.2242 [astro-ph]];\\
D.~Hanson and A.~Lewis,
Phys.\ Rev.\ D {\bf 80}, 063004 (2009)
[arXiv:0908.0963 [astro-ph.CO]];\\
N.~E.~Groeneboom, L.~Ackerman, I.~K.~Wehus and H.~K.~Eriksen,
Astrophys.\ J.\  {\bf 722}, 452 (2010)
[arXiv:0911.0150 [astro-ph.CO]];\\
L.~Ackerman, S.~M.~Carroll and M.~B.~Wise,
Phys.\ Rev.\ D {\bf 75}, 083502 (2007)
[Erratum-ibid.\ D {\bf 80}, 069901 (2009)]
[astro-ph/0701357].

\bibitem{Hanson}
D.~Hanson, A.~Lewis and A.~Challinor,
Phys.\ Rev.\ D {\bf 81}, 103003 (2010)
[arXiv:1003.0198 [astro-ph.CO]].

\bibitem{Watanabe}
M.~a.~Watanabe, S.~Kanno and J.~Soda,
Phys.\ Rev.\ Lett.\  {\bf 102}, 191302 (2009)
[arXiv:0902.2833 [hep-th]].

\bibitem{Gum} 
A.~E.~Gumrukcuoglu, B.~Himmetoglu and M.~Peloso,
Phys.\ Rev.\ D {\bf 81}, 063528 (2010)
[arXiv:1001.4088 [astro-ph.CO]];\\
T.~R.~Dulaney and M.~I.~Gresham,
Phys.\ Rev.\ D {\bf 81}, 103532 (2010)
[arXiv:1001.2301 [astro-ph.CO]].

\bibitem{Watanabe:2010fh}
M.~a.~Watanabe, S.~Kanno and J.~Soda,
Prog.\ Theor.\ Phys.\  {\bf 123}, 1041 (2010)
[arXiv:1003.0056 [astro-ph.CO]].  

\bibitem{Yoko}
S.~Yokoyama and J.~Soda,
JCAP {\bf 0808}, 005 (2008)
[arXiv:0805.4265 [astro-ph]].

\bibitem{Kanno09}
S.~Kanno, J.~Soda and M.~a.~Watanabe,
JCAP {\bf 0912}, 009 (2009)
[arXiv:0908.3509 [astro-ph.CO]].

\bibitem{Rodo}
K.~Dimopoulos, M.~Karciauskas, D.~H.~Lyth and Y.~Rodriguez,
JCAP {\bf 0905}, 013 (2009)
[arXiv:0809.1055 [astro-ph]];\\
M.~Karciauskas, K.~Dimopoulos and D.~H.~Lyth,
Phys.\ Rev.\  D {\bf 80} (2009) 023509
[arXiv:0812.0264 [astro-ph]];\\
C.~A.~Valenzuela-Toledo, Y.~Rodriguez and D.~H.~Lyth,
Phys.\ Rev.\  D {\bf 80}, 103519 (2009)
[arXiv:0909.4064 [astro-ph.CO]];\\
C.~A.~Valenzuela-Toledo and Y.~Rodriguez,
Phys.\ Lett.\  B {\bf 685}, 120 (2010)
[arXiv:0910.4208 [astro-ph.CO]];\\
J.~M.~Wagstaff and K.~Dimopoulos,
Phys.\ Rev.\  D {\bf 83}, 023523 (2011)
[arXiv:1011.2517 [hep-ph]];\\
K.~Dimopoulos,
Int.\ J.\ Mod.\ Phys.\ D {\bf 21}, 1250023 (2012)
[Erratum-ibid.\ D {\bf 21}, 1292003 (2012)]
[arXiv:1107.2779 [hep-ph]].

\bibitem{Riotto}
N.~Bartolo, E.~Dimastrogiovanni, S.~Matarrese and A.~Riotto,
JCAP {\bf 0910}, 015 (2009)
[arXiv:0906.4944 [astro-ph.CO]];\\
N.~Bartolo, E.~Dimastrogiovanni, S.~Matarrese and A.~Riotto,
JCAP {\bf 0911}, 028 (2009)
[arXiv:0909.5621 [astro-ph.CO]];\\
E.~Dimastrogiovanni, N.~Bartolo, S.~Matarrese and A.~Riotto,
Adv.\ Astron.\  {\bf 2010}, 752670 (2010)
[arXiv:1001.4049 [astro-ph.CO]];\\
N.~Bartolo, S.~Matarrese, M.~Peloso and A.~Ricciardone,
arXiv:1306.4160 [astro-ph.CO].

\bibitem{Barnaby} 
N.~Barnaby and M.~Peloso, 
Phys.\ Rev.\ Lett.\  {\bf 106}, 181301 (2011)
[arXiv:1011.1500 [hep-ph]];\\
N.~Barnaby, R.~Namba and M.~Peloso,
JCAP {\bf 1104}, 009 (2011)
[arXiv:1102.4333 [astro-ph.CO]];\\
N.~Barnaby, E.~Pajer and M.~Peloso, 
Phys.\ Rev.\ D {\bf 85}, 023525 (2012)
[arXiv:1110.3327 [astro-ph.CO]];\\
N.~Barnaby, R.~Namba and M.~Peloso,
Phys.\ Rev.\ D {\bf 85}, 123523 (2012)
[arXiv:1202.1469 [astro-ph.CO]].

\bibitem{Ward}
P.~V.~Moniz and J.~Ward, 
Class.\ Quant.\ Grav.\  {\bf 27}, 235009 (2010)
[arXiv:1007.3299 [gr-qc]].

\bibitem{Murata}
K.~Murata and J.~Soda,
JCAP\ {\bf 1106}, 037  (2011)
[arXiv:1103.6164 [hep-th]].

\bibitem{SY}
M.~Shiraishi and S.~Yokoyama,
Prog.\ Theor.\ Phys.\  {\bf 126}, 923 (2011)
[arXiv:1107.0682 [astro-ph.CO]].

\bibitem{Emami}
R.~Emami, H.~Firouzjahi, S.~M.~Sadegh Movahed and M.~Zarei,
JCAP {\bf 1102}, 005 (2011)
[arXiv:1010.5495 [astro-ph.CO]];\\
R.~Emami and H.~Firouzjahi,
JCAP {\bf 1201}, 022 (2012)
[arXiv:1111.1919 [astro-ph.CO]];\\
A.~A.~Abolhasani, R.~Emami, J.~T.~Firouzjaee and H.~Firouzjahi,
JCAP {\bf 1308}, 016 (2013)
[arXiv:1302.6986 [astro-ph.CO]];\\
S.~Baghram, M.~H.~Namjoo and H.~Firouzjahi,
JCAP {\bf 1308}, 048 (2013)
[arXiv:1303.4368 [astro-ph.CO]].

\bibitem{Yamamoto}
K.~Yamamoto, M.~a.~Watanabe and J.~Soda,
Class.\ Quant.\ Grav.\  {\bf 29}, 145008 (2012)
[arXiv:1201.5309 [hep-th]].

\bibitem{Lyth}
M.~Karciauskas,
JCAP {\bf 1201}, 014 (2012)
[arXiv:1104.3629 [astro-ph.CO]];\\
D.~H.~Lyth and M.~Karciauskas,
JCAP {\bf 1305}, 011 (2013)
[arXiv:1302.7304 [astro-ph.CO]].

\bibitem{review}
J.~Soda,
Class.\ Quant.\ Grav.\  {\bf 29}, 083001 (2012)
[arXiv:1201.6434 [hep-th]];\\
A.~Maleknejad, M.~M.~Sheikh-Jabbari and J.~Soda,
Phys.\ Rept.\  {\bf 528}, 161 (2013)
[arXiv:1212.2921 [hep-th]].

\bibitem{solid} 
N.~Bartolo, S.~Matarrese, M.~Peloso and A.~Ricciardone,
JCAP {\bf 1308}, 022 (2013)
[arXiv:1306.4160 [astro-ph.CO]].

\bibitem{Rodriguez} 
Y.~Rodriguez, J.~P.~B.~Almeida and C.~A.~Valenzuela-Toledo,
JCAP {\bf 1304}, 039 (2013)
[arXiv:1301.5843 [astro-ph.CO]];\\
J.~P.~Beltran Almeida, Y.~Rodriguez and C.~A.~Valenzuela-Toledo, 
Mod.\ Phys.\ Lett.\ A {\bf 28}, 1350012 (2013)
[arXiv:1112.6149 [astro-ph.CO]];\\
C.~A.~Valenzuela-Toledo, Y.~Rodriguez and J.~P.~Beltran Almeida,
JCAP {\bf 1110}, 020 (2011)
[arXiv:1107.3186 [astro-ph.CO]].

\bibitem{Bartolo} 
N.~Bartolo, S.~Matarrese, M.~Peloso and A.~Ricciardone,
Phys.\ Rev.\ D {\bf 87}, 023504 (2013)
[arXiv:1210.3257 [astro-ph.CO]].

\bibitem{Shiraishi} 
M.~Shiraishi, E.~Komatsu, M.~Peloso and N.~Barnaby,
JCAP {\bf 1305}, 002 (2013)
[arXiv:1302.3056 [astro-ph.CO]].

\bibitem{Ohashi} 
J.~Ohashi, J.~Soda and S.~Tsujikawa,
Phys.\ Rev.\ D {\bf 87}, 083520 (2013)
[arXiv:1303.7340 [astro-ph.CO]].

\bibitem{Ohashi2} 
J.~Ohashi, J.~Soda and S.~Tsujikawa,
arXiv:1308.4488 [astro-ph.CO].

\bibitem{Ratra} 
B.~Ratra,
Astrophys.\ J.\  {\bf 391}, L1 (1992);\\
J.~Martin and J.~'i.~Yokoyama,
JCAP {\bf 0801}, 025 (2008)
[arXiv:0711.4307 [astro-ph]].

\bibitem{Kanno10} 
S.~Kanno, J.~Soda and M.~-a.~Watanabe,
JCAP {\bf 1012}, 024 (2010)
[arXiv:1010.5307 [hep-th]];\\
S.~Hervik, D.~F.~Mota and M.~Thorsrud,
JHEP {\bf 1111}, 146 (2011)
[arXiv:1109.3456 [gr-qc]];\\
M.~Thorsrud, D.~F.~Mota and S.~Hervik,
JHEP {\bf 1210}, 066 (2012)
[arXiv:1205.6261 [hep-th]].

\bibitem{kinf} 
C.~Armendariz-Picon, T.~Damour and V.~F.~Mukhanov,
Phys.\ Lett.\ B {\bf 458}, 209 (1999)
[hep-th/9904075].

\bibitem{Arkani} 
N.~Arkani-Hamed, H.~-C.~Cheng, M.~A.~Luty and S.~Mukohyama,
JHEP {\bf 0405}, 074 (2004)
[hep-th/0312099].

\bibitem{Piazza} 
F.~Piazza and S.~Tsujikawa,
JCAP {\bf 0407}, 004 (2004)
[hep-th/0405054].

\bibitem{DBI} 
E.~Silverstein and D.~Tong,
Phys.\ Rev.\ D {\bf 70}, 103505 (2004)
 [hep-th/0310221];\\
M.~Alishahiha, E.~Silverstein and D.~Tong,
Phys.\ Rev.\ D {\bf 70}, 123505 (2004)
[hep-th/0404084].

\bibitem{Sami} 
S.~Tsujikawa and M.~Sami,
Phys.\ Lett.\ B {\bf 603}, 113 (2004)
[hep-th/0409212].

\bibitem{Tsuji06} 
S.~Tsujikawa,
Phys.\ Rev.\ D {\bf 73}, 103504 (2006)
[hep-th/0601178];\\
L.~Amendola, M.~Quartin, S.~Tsujikawa and I.~Waga,
Phys.\ Rev.\ D {\bf 74}, 023525 (2006)
[astro-ph/0605488].

\bibitem{Ohashi2011} 
J.~Ohashi and S.~Tsujikawa,
Phys.\ Rev.\ D {\bf 83}, 103522 (2011)
[arXiv:1104.1565 [astro-ph.CO]].

\bibitem{Unnikrishnan:2013vga} 
S.~Unnikrishnan and V.~Sahni,
JCAP {\bf 1310}, 063 (2013)
[arXiv:1305.5260 [astro-ph.CO]].

\bibitem{Kao}
T.~Q.~Do and W.~F.~Kao,
Phys.\ Rev.\  D {\bf 84}, 123009 (2011);\\
T.~Q.~Do, W.~F.~Kao and I.~-C.~Lin,
Phys.\ Rev.\ D\ {\bf 83}, 123002  (2011).


\bibitem{Tachyon}
S.~Bhowmick and S.~Mukherji,
Mod.\ Phys.\ Lett.\ A {\bf 27}, 1250009 (2012)
[arXiv:1105.4455 [hep-th]].

\bibitem{expo} 
J.~J.~Halliwell,
Phys.\ Lett.\  B {\bf 185}, 341 (1987);\\
F.~Lucchin and S.~Matarrese,
Phys.\ Rev.\  D {\bf 32}, 1316 (1985);\\
J.~Yokoyama and K.~i.~Maeda,
Phys.\ Lett.\  B {\bf 207}, 31 (1988).

\bibitem{Garriga} 
J.~Garriga and V.~F.~Mukhanov,
Phys.\ Lett.\ B {\bf 458}, 219 (1999)
[hep-th/9904176].

\bibitem{Carroll} 
S.~M.~Carroll, M.~Hoffman and M.~Trodden,
Phys.\ Rev.\ D {\bf 68}, 023509 (2003)
[astro-ph/0301273].

\bibitem{Wald} 
R.~M.~Wald,
Phys.\ Rev.\ D {\bf 28}, 2118 (1983).

\bibitem{nongau} 
D.~Seery and J.~E.~Lidsey,
JCAP {\bf 0506}, 003 (2005)
[astro-ph/0503692];\\
X.~Chen, M.~-x.~Huang, S.~Kachru and G.~Shiu,
JCAP {\bf 0701}, 002 (2007)
[hep-th/0605045].

\end{thebibliography}
\end{document}